# Two improved normalized subband adaptive filter algorithms with good robustness against impulsive interferences

Yi Yu[1, 2] • Haiquan Zhao[1, 2] • Badong Chen[3] • Zhengyou He[2]

**Abstract.** To improve the robustness of subband adaptive filter (SAF) against impulsive interferences, we propose two modified SAF algorithms with an individual scale function for each subband, which are derived by maximizing correntropy-based cost function and minimizing logarithm-based cost function, respectively, called MCC-SAF and LC-SAF. Whenever the impulsive interference happens, the subband scale functions can sharply drop the step size, which eliminate the influence of outliers on the tap-weight vector update. Therefore, the proposed algorithms are robust against impulsive interferences, and exhibit the faster convergence rate and better tracking capability than the sign SAF (SSAF) algorithm. Besides, in impulse-free interference environments, the proposed algorithms achieve similar convergence performance as the normalized SAF (NSAF) algorithm. Simulation results have demonstrated the performance of our proposed algorithms.

## 1  Introduction

Adaptive filtering algorithms have found a large amount of applications such as system identification, acoustic echo cancellation (AEC) and channel equalization, etc [4], [10], [18]. It is well-known that the least mean square (LMS) algorithm and its normalized version (NLMS) are very popular due to their simplicity. However, its convergence speed continues to be unsatisfactory for colored input signals. In order to speed up the convergence in practical applications that entail colored input signals, an

Haiquan Zhao (✉)
hqzhao@home.swjtu.edu.cn

Yi Yu
yuyi_xyuan@163.com

Badong Chen
chenbd@mail.xjtu.edu.cn

[1] Key Laboratory of Magnetic Suspension Technology and Maglev Vehicle, Ministry of Education, Chengdu, China.
[2] School of Electrical Engineering, Southwest Jiaotong University, Chengdu, China
[3] School of Electronic and Information Engineering, Xi'an Jiaotong University, Xi'an, China.

attractive approach is to use the subband adaptive filter (SAF), because it partitions the colored input signal into nearly white subband signals [4], [23]. In [5], Lee and Gan developed the normalized SAF (NSAF) algorithm, which provides faster convergence rate and almost the same computational complexity as compared to the NSAF. Afterwards, to overcome a compromise of the NSAF between fast convergence rate and low steady-state error, several variable step size NSAF algorithms were proposed [2], [8], [11], [12], [19], [20]. Regrettably, most of the aforementioned algorithms were derived by solving the $L_2$-norm-based optimization problem, thus their convergence performances are seriously damaged by impulsive interferences (which are often encountered in practical applications).

It has been shown that some adaptive filtering algorithms that minimize the $L_1$-norm of the error signal offer strong anti-jamming capability to impulsive interferences [6], [14]. In [7], Ni *et al* proposed a sign SAF (SSAF) algorithm by incorporating the $L_1$-norm optimization into the subband filter, which exhibits good robustness against impulsive interferences and fast convergence for colored input signals, and also proposed a variable regularization parameter SSAF (VRP-SSAF) algorithm to further reduce the steady-state error. Following these works, many researchers have developed various variants to further improve the performance of the SSAF algorithm, such as variable step size SSAF algorithms [3], [13], [22], affine projection SSAF (AP-SSAF) algorithm [9] and proportionate SSAF (P-SSAF) algorithm [9].

Recently, both the maximum correntropy criterion (MCC) [1], [15] and minimum logarithm criterion (LC) [16], [17] were applied to the LMS-type filtering algorithms, respectively, by using the gradient rule. However, in my opinion, these two strategies reflect a common phenomenon in the adaptation of the resulting algorithms, even if their principles are different. Namely, the step size will immediately become very small as long as the impulsive interference occurs. This is the reason why these MCC- and LC-based LMS algorithms possess the robustness to impulsive interferences. Therefore, this paper incorporates these two criteria into the SAF to deal with the impulsive interferences, respectively, and thus derives MCC-SAF and LC-SAF algorithms. Both proposed algorithms have the following features:
1) Both proposed algorithms use an individual scale function for each subband, which instantly shrinks the step size whenever the impulsive interference happens. This leads to the robustness of the proposed algorithms against impulsive interferences.
2) Compared with the SSAF and AP-SSAF algorithms, both proposed algorithms have faster convergence rate and better tracking capability.
3) Both proposed algorithms reach the convergence performance similar to the NSAF in impulse-free interference environments.

## 2  Review the NSAF Algorithm

Consider the desired signal $d(n)$ that is the output signal of the system

$$d(n) = \mathbf{u}^T(n)\mathbf{w}_o + \eta(n),  \tag{1}$$

where the superscript $T$ denotes transposition, $\mathbf{w}_o$ is an unknown $M$-dimension impulse response that needs to be identified, $\mathbf{u}(n) = [u(n), u(n-1), ..., u(n-M+1)]^T$ is the input signal vector, $\eta(n) = \upsilon(n) + \theta(n)$ is the additive noise including the measurement noise $\upsilon(n)$ and impulsive interference $\theta(n)$. Fig. 1 shows the multiband structure of SAF [5], where $N$ denotes number of subbands. The desired signal $d(n)$ and the input signal $u(n)$ are divided into $N$ subband signals by using analysis filters $\{H_i(z), i \in [0, N-1]\}$, respectively. Then, the subband signals $y_i(n)$ and $d_i(n)$ for $i \in [0, N-1]$ are critically decimated to yield $y_{i,D}(k)$ and $d_{i,D}(k)$, respectively, where $n$ and $k$ are used to indicate the original sequences and the decimated sequences. The $i$th subband output signal is described by $y_{i,D}(k) = \mathbf{u}_i^T(k)\mathbf{w}(k)$, where $\mathbf{w}(k)$ is the tap-weight vector of adaptive filter, and $\mathbf{u}_i(k) = [u_i(kN), u_i(kN-1), ..., u_i(kN-M+1)]^T$. The output error of the $i$th subband is defined as

$$e_{i,D}(k) = d_{i,D}(k) - y_{i,D}(k) = d_{i,D}(k) - \mathbf{u}_i^T(k)\mathbf{w}(k) \qquad (2)$$

where $d_{i,D}(k) = d_i(kN)$.

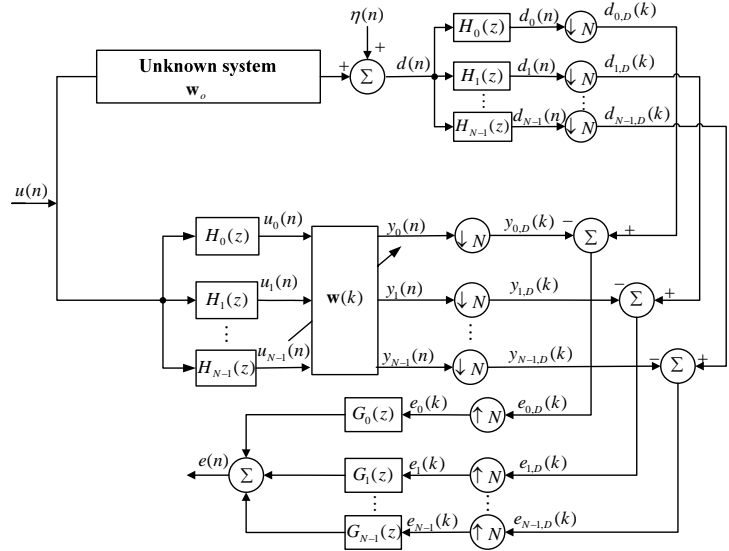

**Fig. 1.** Structure of subband adaptive filter

The conventional NSAF algorithm can be derived by minimizing the cost function, defined by [5]

$$J(k) = \frac{1}{2} \sum_{i=0}^{N-1} \frac{e_{i,D}^2(k)}{\|\mathbf{u}_i(k)\|^2} \qquad (3)$$

where $\|\cdot\|$ denotes the $L_2$-norm of vector. Based on the gradient descent theory, the tap-weight vector of the NSAF is updated as

$$\mathbf{w}(k+1) = \mathbf{w}(k) + \mu \sum_{i=0}^{N-1} \frac{e_{i,\mathrm{D}}(k)\mathbf{u}_i(k)}{\|\mathbf{u}_i(k)\|^2} \qquad (4)$$

where $\mu$ is the step size.

## 3  Proposed Algorithms

To improve the robustness of SAF combating impulsive interferences, in this section, the MCC-SAF and LC-SAF algorithms are derived.

### 3.1  MCC-SAF algorithm

As described in [1], [15], the correntropy can measure local similarity between two arbitrary random variables X and Y, which is defined as follows:

$$V(X, Y) = E[\kappa(X, Y)] = \int \kappa(x, y) dF_{X,Y}(x, y) \qquad (5)$$

where $\kappa(x, y)$ is a shift-invariant Mercer kernel, and $F_{X,Y}(x, y)$ denotes the joint distribution function of $(X, Y)$. In the correntropy, the most widely used kernel is the Gaussian kernel, i.e.,

$$\kappa(x, y) = \frac{1}{\sqrt{2\pi}\sigma} \exp(-\frac{e^2}{2\sigma^2}) \qquad (6)$$

where $\sigma > 0$ is the kernel width, and $e = x - y$. Given $x = d_{i,\mathrm{D}}(k)$ and $y = y_{i,\mathrm{D}}(k)$, we can obtain the following correntropy-based cost function to derive the MCC-SAF, i.e.,

$$J_{\mathrm{MCC}}(k) = \frac{1}{2\beta} \sum_{i=0}^{N-1} \exp\left[-\beta \frac{e_{i,\mathrm{D}}^2(k)}{\|\mathbf{u}_i(k)\|^2}\right] \qquad (7)$$

where $\beta$ is called the kernel parameter associated with the kernel width $\sigma$ in (6) by $\beta = 1/2\sigma^2$. Taking the gradient of $J_{\mathrm{MCC}}(k)$ with respect to $\mathbf{w}(k)$, we have

$$\nabla J_{\mathrm{MCC}}(k) = \sum_{i=0}^{N-1} \exp\left[-\beta \frac{e_{i,\mathrm{D}}^2(k)}{\|\mathbf{u}_i(k)\|^2}\right] \times \frac{e_{i,\mathrm{D}}(k)\mathbf{u}_i(k)}{\|\mathbf{u}_i(k)\|^2}. \qquad (8)$$

By using the gradient ascent approach that maximizes the cost function $J_{\mathrm{MCC}}(k)$, the tap-weight update of the proposed MCC-SAF algorithm is expressed as

$$\begin{aligned}\mathbf{w}(k+1) &= \mathbf{w}(k) + \mu \nabla J_{\mathrm{MCC}}(k) \\ &= \mathbf{w}(k) + \mu \sum_{i=0}^{N-1} \exp\left[-\beta \frac{e_{i,\mathrm{D}}^2(k)}{\|\mathbf{u}_i(k)\|^2}\right] \times \frac{e_{i,\mathrm{D}}(k)\mathbf{u}_i(k)}{\|\mathbf{u}_i(k)\|^2}.\end{aligned} \qquad (9)$$

### 3.2 LC-SAF algorithm

Motivated by the robustness of the logarithmic-based cost function against impulsive interferences [16], [17], we introduce it into the SAF, and then obtaining a new cost function

$$J_{LC}(k) = \frac{1}{2\gamma} \sum_{i=0}^{N-1} \ln\left[1 + \gamma \frac{e_{i,D}^2(k)}{\|\mathbf{u}_i(k)\|^2}\right] \quad (10)$$

where $\gamma > 0$ is a flexible parameter. Note that when $N = 1$, (10) has been reported in [16] to derive the NLMS-type algorithm.

Taking the gradient of $J_{LC}(k)$ with respect to $\mathbf{w}(k)$, we achieve

$$\nabla J_{LC}(k) = -\sum_{i=0}^{N-1} \frac{1}{1 + \gamma\left(e_{i,D}(k)/\|\mathbf{u}_i(k)\|\right)^2} \times \frac{e_{i,D}(k)\mathbf{u}_i(k)}{\|\mathbf{u}_i(k)\|^2}. \quad (11)$$

Similar to the previous described NSAF, we obtain the tap-weight update of the LC-SAF algorithm via using the gradient descent approach,

$$\mathbf{w}(k+1) = \mathbf{w}(k) - \mu \nabla J_{LC}(k)$$

$$= \mathbf{w}(k) + \mu \sum_{i=0}^{N-1} \frac{1}{1 + \gamma\left(e_{i,D}(k)/\|\mathbf{u}_i(k)\|\right)^2} \times \frac{e_{i,D}(k)\mathbf{u}_i(k)}{\|\mathbf{u}_i(k)\|^2}. \quad (12)$$

### 3.3 Discussions

***Remark 1:*** Without loss of generality, the proposed MCC-SAF and LC-SAF algorithms, i.e., (9) and (12), can be described in an unified form as

$$\mathbf{w}(k+1) = \mathbf{w}(k) + \mu \sum_{i=0}^{N-1} f_i(\bullet) \frac{e_{i,D}(k)\mathbf{u}_i(k)}{\|\mathbf{u}_i(k)\|^2} \quad (13)$$

where $f_i(\bullet)$ denotes the scale function of the $i$th subband with regard to the corresponding normalized error $e_{i,D}(k)/\|\mathbf{u}_i(k)\|$. In the MCC-SAF, the subband scale functions $f_i(\bullet)$ for $i \in [0, N-1]$ are given by

$$f_{MCC,i}\left(\frac{e_{i,D}(k)}{\|\mathbf{u}_i(k)\|}\right) = \exp\left[-\beta\left(\frac{e_{i,D}(k)}{\|\mathbf{u}_i(k)\|}\right)^2\right], \quad (14)$$

and the subband scale functions of the MCC-SAF are expressed as

$$f_{LC,i}\left(\frac{e_{i,D}(k)}{\|\mathbf{u}_i(k)\|}\right) = \frac{1}{1 + \gamma\left(e_{i,D}(k)/\|\mathbf{u}_i(k)\|\right)^2}. \quad (15)$$

Evidently, for the standard NSAF algorithm, the scale function for each subband is a constant 1, i.e.,

$$f_{NSAF,i}(\bullet) = 1 \quad \text{for} \quad i \in [0, N-1]. \quad (16)$$

Hence, we can conclude that both the proposed MCC-SAF and LC-SAF algorithms can be considered as the NSAF algorithm with particular subband scale functions. Moreover, (14) and (15) will become (16) when the parameters $\beta$ and $\gamma$ tend to zero, thereby both the proposed MCC-SAF and LC-SAF algorithms will be reduced to the NSAF. However, two proposed algorithms have good robustness against impulsive interferences, see Remark 2.

***Remark 2:*** To visually see the property of the proposed algorithms, Fig. 2 shows the shapes of the subband scale functions given by (14) and (15) with respect to the normalized subband error $e_{i,\text{D}}(k)/\|\mathbf{u}_i(k)\|$, respectively, using the various values of $\beta$ and $\gamma$. Cleary, no matter when the impulsive interference happens, $e_{i,\text{D}}(k)/\|\mathbf{u}_i(k)\|$ has very large magnitude, and then the corresponding scale function dramatically reduces the step size to a very small value. It was found in [16] that a small step size can suppress the influence of wrong information caused by outliers such as the impulsive interferences on the update of the tap-weight vector. Therefore, the proposed algorithms provide good robustness against impulsive interferences by shrinking the step size. On the other hand, if there is no impulsive interference, the magnitude of $e_{i,\text{D}}(k)/\|\mathbf{u}_i(k)\|$ will be very small. In this case, the proposed algorithms behave like the NSAF, since the subband scale functions are approximately equal to 1. In other words, the subband scale functions shown in (14) and (15) only work in the case of the appearance of impulsive interferences. More importantly, the proposed subband scale functions, i.e., (14) and (15), can be directly applied to the existing variable step size NSAF algorithms (e.g., [2], [8], [11], [12], [19]) to improve their robustness against impulsive interferences.

***Remark 3:*** It has been found that the NSAF is stable for convergence as long as the step size satisfies

$$0 < \mu < 2 . \tag{17}$$

And according to Remark 1 and 2, there are always $f_{\text{MCC},i}(\bullet) \leq 1$ and $f_{\text{LC},i}(\bullet) \leq 1$ in the MCC-SAF and LC-SAF algorithms, respectively. Thus, (17) is also the convergence condition of both proposed algorithms.

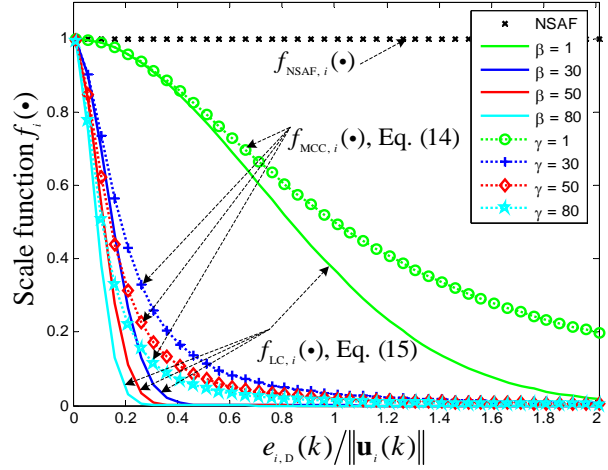

**Fig. 2.** Scale function of the $i$th subband.

*Remark 4:* Table 1 provides the computational complexities of both proposed algorithms and some existing SAF algorithms in terms of the total number of multiplications, additions, divisions, and exponents for each input sample. Compared with the NSAF algorithm, the additional cost of both proposed MCC-SAF and LC-SAF algorithms stem from the computation of (14) and (15), respectively. Since $e_{i,\text{D}}(k)/\|\mathbf{u}_i(k)\|^2$ computation is already available from (13), the MCC-SAF algorithm only requires an additional 3 multiplications and 1 exponent for each fullband input sample, and the LC-SAF algorithm requires an additional 3 multiplications, 1 division and 1 addition. For the MCC-SAF algorithm, the exponent calculation can be implemented by the form of table look-at to alleviate the complexity at the expense of memory space. So, both proposed algorithms are almost the same in the computation complexity. Importantly, the slight increase of these two proposed algorithms in computational complexity in comparison with the NSAF can be compromised by their robustness against impulsive interferences.

**Table** 1. Computational complexity of various SAF algorithms for each input sample. The integer $L$ denotes the length of the prototype filter of the filter bank, and $P$ denotes the affine projection order.

| Algorithms | Multiplications | Additions | Divisions | exponent |
|---|---|---|---|---|
| NSAF | $3M+3NL+1$ | $3M+3N(L-1)$ | 1 | 0 |
| SSAF | $M+2M/N+3NL$ | $2M+M/N+3NL-2N-1$ | $1/N$ | 0 |
| AP-SSAF | $PM+2M/N+3NL$ | $2PM+M/N+3NL-2N-1$ | $1/N$ | 0 |
| LC-SAF | $3M+3NL+4$ | $3M+3N(L-1)+1$ | 2 | 0 |
| MCC-SAF | $3M+3NL+4$ | $3M+3N(L-1)$ | 1 | 1 |

## 4  Simulations

In this section, the proposed MCC-SAF and LC-SAF algorithms are evaluated using Monte Carlo (MC) simulations (average of 50 independent runs). The unknown $\mathbf{w}_o$ is a realistic acoustic impulse response with $M = 512$ taps. In our simulations, the measurement noise $\upsilon(n)$ is white Gaussian noise, with a signal-to-noise rate (SNR) of 30dB; and a four-subband (i.e., $N = 4$) cosine modulated filter bank is used, with 60 dB stop-band attenuation, whose prototype has 32 tap-weights [4], [21]. The normalized mean square deviation (NMSD), $10\log_{10}(\|\mathbf{w}_o - \mathbf{w}(k)\|_2^2 / \|\mathbf{w}_o\|_2^2)$, dB, is used as a measure of the algorithm performance.

### 4.1  System identification

In this scenario, the colored input signal $u(n)$ is generated by filtering a zero-mean white Gaussian signal through a first-order autoregressive (AR(1)) system $G(z) = 1/(1 - 0.9z^{-1})$ [3], [13]. The impulsive interference $\theta(n)$ is usually modeled as a Bernoulli-Gaussian (BG) process, i.e., $\theta(n) = c(n)A(n)$ [3], [13], where $c(n)$ is a Bernoulli process with the probability mass function described by $p\{c(n) = 1\} = P_r$ and $p\{c(n) = 0\} = 1 - P_r$ (with $P_r$ being the probability of the occurrence of the impulsive interference), and $A(n)$ is a white Gaussian process with zero mean and variance $\sigma_A^2 = 100E\left[\left(\mathbf{u}^T(n)\mathbf{w}_o\right)^2\right]$.

*1) Effect of $\beta$ and/or $\gamma$*

Fig. 3 shows the steady-state NMSDs of the MCC-SAF and LC-SAF algorithms versus the parameters $\beta$ and $\gamma$, respectively, in impulsive interference environments with $P_r = 0.001$, 0.005, 0.01 and 0.05. Here, the step size for both algorithms is selected to be the same value, i.e., $\mu = 0.1$ or $\mu = 0.5$; and the results are obtained by time-averaging 500 instantaneous NMSDs in the steady-state. Clearly, both algorithms have larger steady-state NMSD for larger values of $P_r$. Values of $\beta$ and/or $\gamma$ are higher, the steady-state NMSDs of both algorithms are smaller, since the capabilities of the subband scale functions in (14) and (15) to shrink the step size are stronger (also see Fig. 2). However, their values can not be too large, because we also need to consider the convergence rate of gradient-based adaptive methods. In addition, using the same step size and $\beta = \gamma$, the steady-state NMSD of the LC-SAF algorithm is not as low as that of the MCC-SAF algorithm.

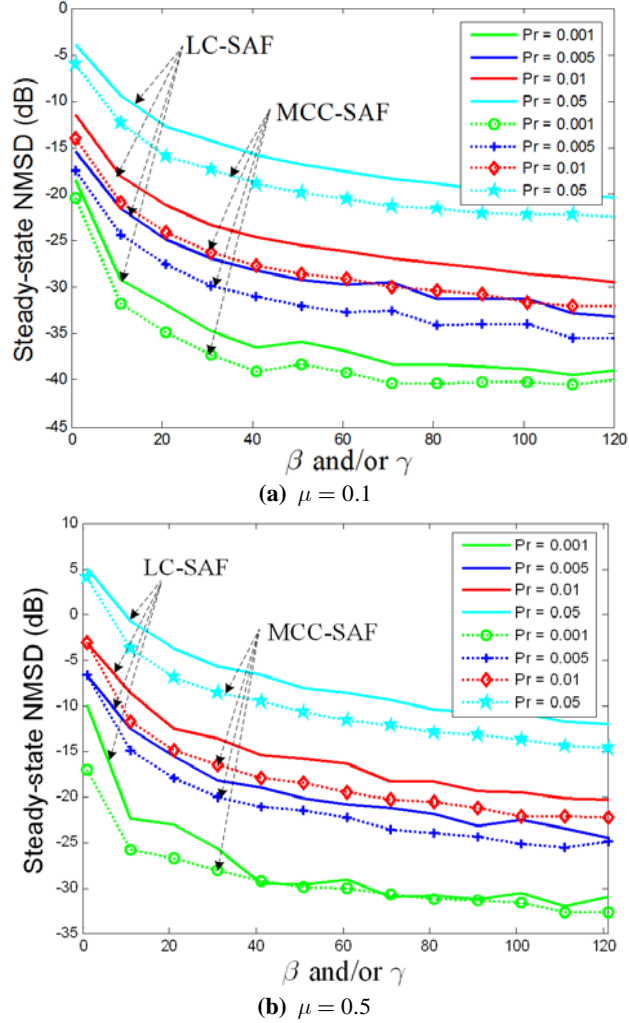

**Fig. 3.** Steady-state NMSDs of the MCC-SAF and LC-SAF algorithms using different values of $\beta$ and $\gamma$, respectively.

*2) No impulsive interference*

This example compares the performance of the proposed algorithms with that of the NSAF, SSAF and AP-SSAF algorithms, where there is no impulsive interference (i.e., $P_r = 0$), as shown in Fig. 4. To assess the tracking capability of the algorithms, the unknown $\mathbf{w}_o$ changes as $-\mathbf{w}_o$ at the $8 \times 10^4$ th input samples. To obtain a fair comparison, we select values of parameters in such a way that all algorithms reach the same steady-state NMSD. The affine projection order in the AP-SSAF is chosen as $P = 4$, since its computational complexity increases as $P$ increases. It is clear that the performance of SSAF algorithm is the worst in convergence rate in the absence of impulsive interferences, due to the fact that it only uses the sign information of subband errors to update the tap-weight vector. The AP-SSAF algorithm can improve convergence performance, but it sacrifices computation cost. Interestingly, the proposed algorithms are slightly slower than the NSAF algorithm in terms of convergence rate and tracking capability. This reason is that the normalized subband errors $e_{i,\,\mathrm{D}}(k)\big/\|\mathbf{u}_{i,\,\mathrm{D}}(k)\|$ will be very small when the impulsive interference does not appear, and thus the subband scale functions are approximately equal to a constant 1.

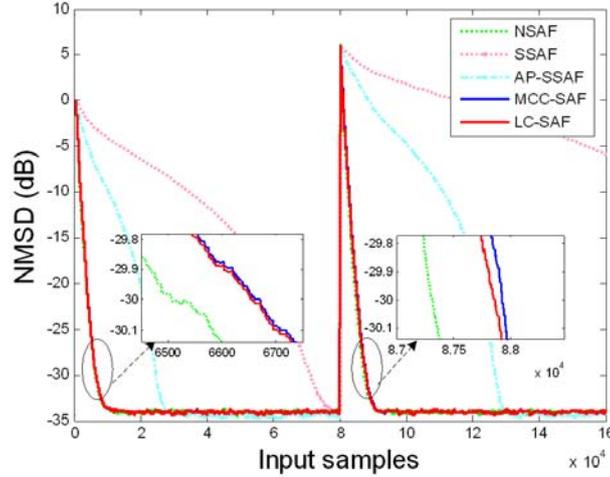

**Fig. 4.** The NMSD curves of various SAF algorithms for AR(1) input, in the absence of impulsive interference. NSAF: $\mu = 0.65$; SSAF: $\mu = 0.004$; AP-SSAF: $\mu = 0.0085$, $P = 4$; MCC-SAF: $\mu = 0.65$, $\beta = 10$; LC-SAF: $\mu = 0.65$, $\gamma = 10$.

*3) Under impulsive interference*

In Figs. 5 and 6, the NMSD performances of these algorithms are compared in the presence of impulsive interferences, where $P_r = 0.005$ for Fig. 5 and $P_r = 0.01$ for Fig. 6. Among these algorithms, only the NSAF algorithm based on the $L_2$-norm is divergent. Although the SSAF and AP-SSAF algorithms derived from the $L_1$-norm optimization are robust to impulsive interferences, their weaknesses are slow convergence and also the AP-SSAF requires high computation. Interestingly, the proposed MCC-SAF and LC-SAF algorithms work well in impulsive interference environments, and can faster converge than the SSAF and AP-SSAF as well as track the change of the unknown system (e.g., at the $5\times10^4$ th input samples). This is because that the subband scale functions, shown by (14) and/or (15), can promptly reduce the step size as long as the impulsive interference appears.

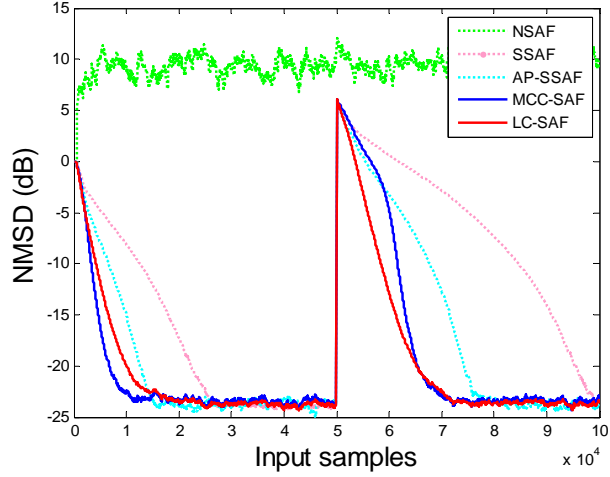

**Fig. 5.** The NMSD curves of various SAF algorithms for AR(1) input, with impulsive interference $P_r = 0.005$. NSAF: $\mu = 0.25$; SSAF: $\mu = 0.015$; AP-SSAF: $\mu = 0.018$, $P = 4$; MCC-SAF: $\mu = 0.45$, $\beta = 80$; LC-SAF: $\mu = 0.25$, $\gamma = 80$.

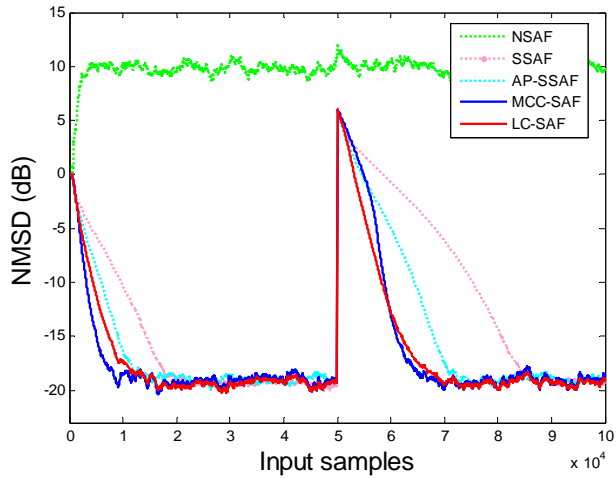

**Fig. 6.** The NMSD curves of various SAF algorithms for AR(1) input, with impulsive interference $P_r = 0.01$. Parameters' setting is the same as Fig. 5.

### 4.2 Acoustic echo cancellation with double-talk

In this example, we examine the performances of both proposed algorithms in an acoustic echo cancellation application with double-talk, as shown in Fig. 7. The main goal of echo cancellation is to identify the echo path $\mathbf{w}_o$, but the far-end input signal $u(n)$ is a speech signal. Besides, the near-end signal that can be considered as the impulsive interference $\theta(n)$ is a speech signal. As one can see, all algorithms except the NSAF are robust against double-talk happened in the period with sample index of $[2, 4] \times 10^4$. Moreover, both proposed algorithms have better performance than the SSAF and AP-SSAF algorithms.

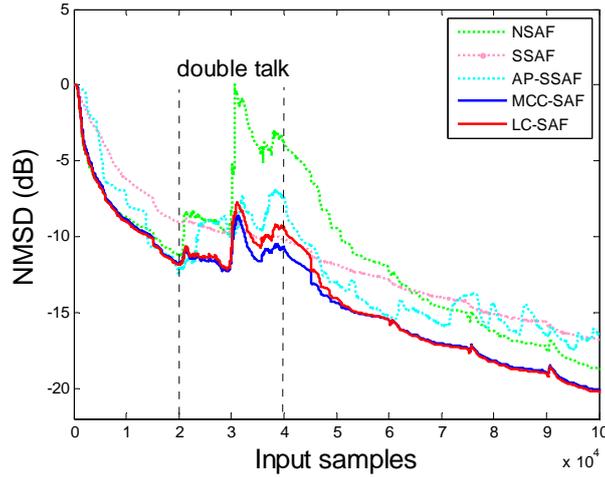

**Fig.** 7. The NMSD curves of various SAF algorithms in acoustic echo cancellation with double talk. NSAF: $\mu = 0.35$. The parameters of other algorithms are the same as Fig. 5.

## 5    Conclusions

In this study, the MCC-SAF and LC-SAF algorithms have been introduced by maximizing the correntropy-based cost function and minimizing the logarithm-based cost function, respectively. In the tap-weight vector update, each subband receives an individual scale function, which instantly shrinks the step size whenever the impulsive interference appears. This eliminates the influence of outliers (e.g., impulsive interferences) on the convergence performance of the proposed algorithms. Simulation results have shown that the proposed algorithms outperform the SSAF and AP-SSAF algorithms in terms of convergence rate and tracking capability in impulsive interference situations. Furthermore, their convergence performance is comparable to the NSAF algorithm in the absence of impulsive interferences. The performance analysis of both proposed algorithms is our future work.

### Acknowledgments

This work was partially supported by National Science Foundation of P.R. China (Grant: 61271340, 61571374, 61134002, 61433011 and U1234203), the Sichuan Provincial Youth Science and Technology Fund (Grant: 2012JQ0046), and the Fundamental Research Funds for the Central Universities (Grant: SWJTU12CX026).